\documentclass{xray_symp}
\usepackage{graphics}

\begin{document}
\title{Large-scale structures in the distribution of
       X-ray selected AGN}
\author{D. Engels\inst{1}, F. Tesch\inst{1},
        C. Ledoux\inst{2}, J. Wei\inst{3}, A. Ugryumov\inst{4},
        D. Valls-Gabaud\inst{2}, J. Hu\inst{3},  \and W. Voges\inst{5}}  
\institute{Hamburger Sternwarte, Gojenbergsweg 112, D-21029 Hamburg, Germany
\and Observatoire Astronomique de Strasbourg, 11 Rue de l'Universit\'{e}, 
                                                F-67000 Strasbourg, France
\and Beijing Astronomical Observatory, Chinese Academy of Sciences, 
       Beijing 100080, P.R. China 
\and Special Astrophysical Observatory, RAS, 
Nizhnij Arkhyz, Karachai-Circessia, Russia
\and Max--Planck--Institut f\"ur extraterrestrische Physik,  
           Giessenbachstra{\ss}e, D-85740 Garching, Germany
}
\authorrunning{Engels et al.}
\titlerunning{Large-scale structures in the distribution of
       X-ray selected AGN}
\maketitle

\begin{abstract}

We are searching for large-scale structures in the distribution of AGN 
discovered by the ROSAT All-Sky Survey. The RASS detected 
$>$ 60000 X-ray objects, of which $\approx$ 35\% are AGN at z $<$ 0.5.  
The surface density in the extragalactic sky is $\approx$ 0.5 
AGN/deg$^{2}$, which has not been reached until now by any other survey for 
this redshift range. We efficiently single out the AGN among all the
RASS sources using the Hamburg/RASS database of optical identifications,
which contains presently 13867 entries.

Follow-up spectroscopy of RASS AGN candidates identified in selected
areas of the northern sky is underway to determine the spatial distribution
of the AGN in these areas. In every area structures reminiscent of clusters
and filaments are found on scales 50 - 100 h$^{-1}$Mpc. These structures
have a similar size as the serendipitously discovered groups of AGN in optical
surveys at higher redshifts.
Samples of low redshift AGN drawn from the RASS are 
large enough to be used with samples of higher redshift to study evolutionary
effects in the large-scale distribution of AGN. 

\end{abstract}

\section{Introduction}

Wideangle galaxy surveys have convincingly shown
 the existence of large-scale structure
in the matter distribution of the universe in form of filaments around voids 
with extensions\footnote{Sizes are given in h$^{-1}$Mpc 
with h = H$_0$/100 and q$_0$=0.5. H$_0$ is the Hubble constant}
 of 20 - 30 h$^{-1}$Mpc and in form of  superstructures like 
the ``Great Wall'' with an  extension of 60 $\times$ 170 h$^{-1}$Mpc 
(Geller \& Huchra 1989). 
Such structures have been found up to a distance 
of z $<$ 0.2 studying mostly clusters of galaxies. Recently Einasto et al. 
(1997a,b) claimed that clusters of galaxies form 
nodes and walls in a rather 
regular supercluster void network with a typical scale of 120 h$^{-1}$Mpc. 
To study such structures at even higher redshifts needs the involvement of 
quasars, either by using their distribution itself or the distribution of 
absorbers in the line of sight (e.g. Quashnock et al. 1997).

Using different samples of AGN (=Sy1 galaxies and QSOs) indications for 
structures on scales $<$ 20 h$^{-1}$Mpc have been found, but not on larger 
ones (Croom \& Shanks 1996 and ref. therein). 
Attempts to study the evolution
of these structures lead to divergent results. Iovino \& Shaver (1988) found 
increasing strength of clustering with decreasing redshift, while La Franca
et al. (1998) obtained the opposite result. Shanks \& Boyle (1994) obtained
a result compatible with no evolution. All these evolutionary
studies are hampered by the lack of suitable low-redshift AGN samples,
as these require wide-angle (better: all-sky) surveys. 
Evidence for structures $>$ 20 h$^{-1}$Mpc comes from the serendipituos 
discovery of clusters of quasars with extensions of 60 - 200 h$^{-1}$Mpc
(Graham et al. 1995 and ref. therein). Komberg et al. (1996) identified 
12 Large Quasar Groups of similar size in the V\'{e}ron-Cetty \& V\'{e}ron QSO 
catalogue at redshifts 0.5 $<$ z $<$ 2.0, the lower redshift limit being 
probably due to the lack of suitable wide-angle surveys.

The ROSAT All-Sky Survey with its tens of thousands of AGN offers
an excellent possibility to investigate the distribution of AGN in the
local universe and to search for close clusters of AGN. If AGN clusters
at low redshifts (z$\approx$0.1) can be found, a comparison with the
distribution of galaxies in the cluster region will be feasible. This
would allow us 
to relate the observed AGN clustering directly with clustering of
galaxies.

\begin{figure}
\resizebox{\hsize}{!}{\includegraphics{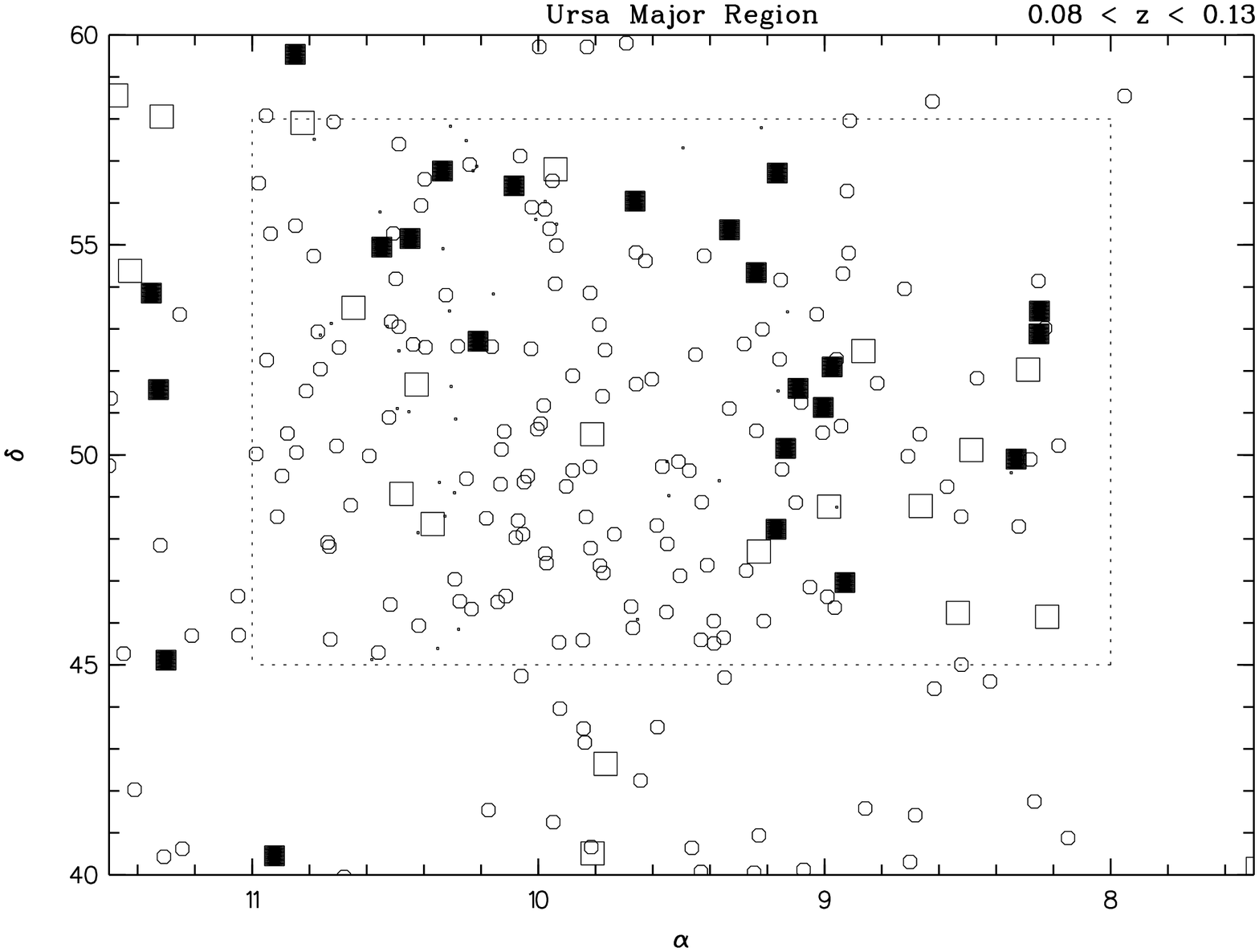}}
\caption[]{AGN distribution in the selected area Ursa Major. A 
filamentary structure is outlined by the filled squares.
AGN in front of (large squares) and behind the structure (small squares), and 
AGN with unknown redshift (dots) are coded respectively. The dotted lines 
enclose the area chosen for complete spectroscopy. }
\resizebox{\hsize}{!}{\includegraphics{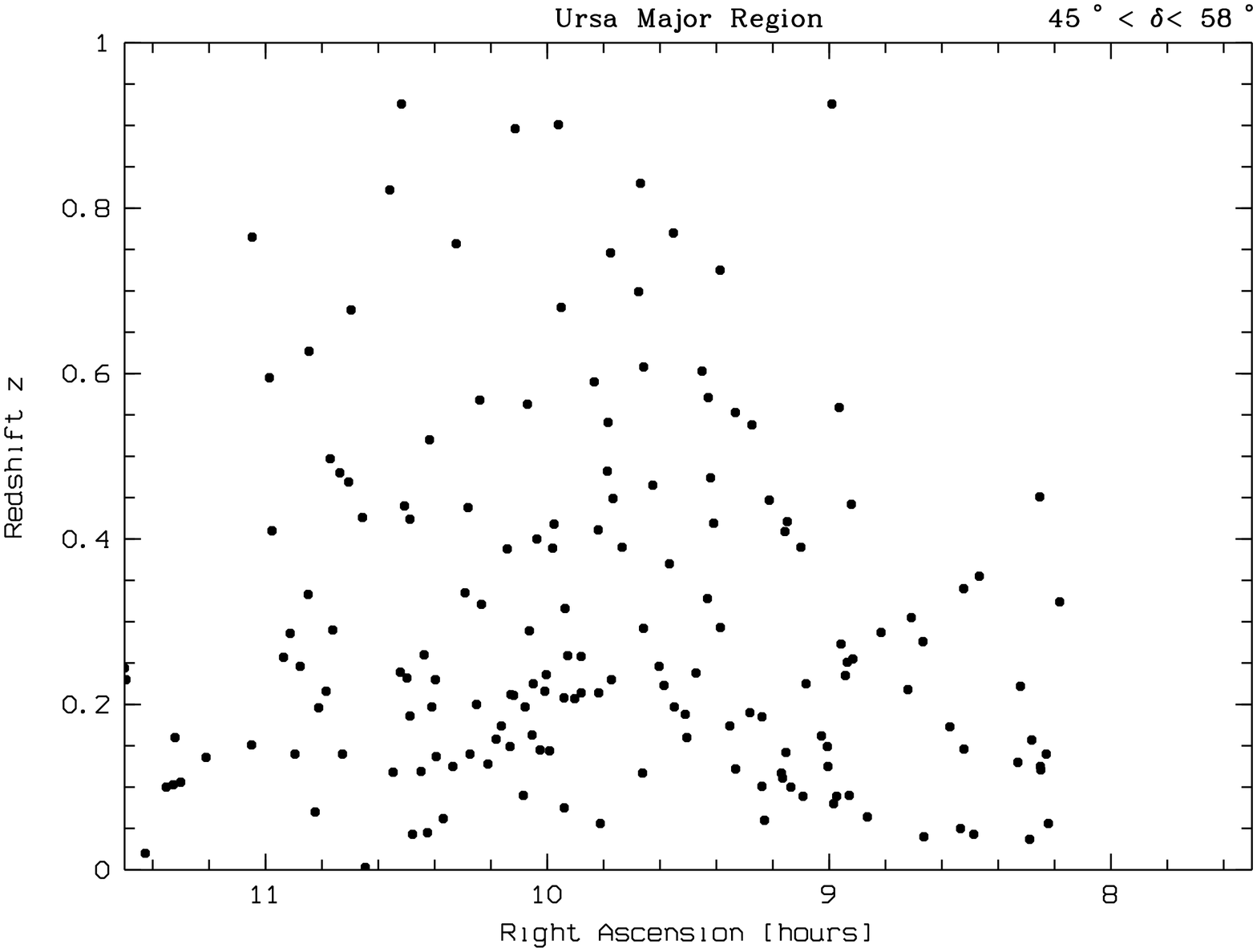}}
\caption[]{Selected area Ursa Major. Slice merging all AGN
with 45$<\delta <$58$^{\circ}$ in the right ascension - redshift plane.}
\end{figure}

\section{The Hamburg/RASS database of optical identifications}
The redshift determination for large samples of RASS selected AGN would 
require unacceptable large amounts of telescope time, if identification
of the correct optical counterpart would be required simultaneously.
We are however able to single out the AGN from the RASS sources very
efficiently by cross-correlating the RASS with the optical data from
the Hamburg Quasar Survey (HQS; Hagen et al. 1995). RASS sources are
identified by Bade et al. (1998)
systematically on the digitized objective prism plates  
of the HQS and their
spectroscopic follow-up observations gave a satisfying high 
confirmation rate $\ge$90\% (Bade et al. 1992) for the AGN candidates.
The Hamburg/RASS database presently contains 13867 entries, covering
RASS sources selected from $>$10\,000\,deg$^2$ of the extragalactic
northern sky. The database is regularly
updated and enlarged with information from the newest plates taken for the
HQS and from subsequent processings of the RASS. Optical identifications
are available for 61\% of the RASS sources processed, with 50\% being
AGN, 40\% stars and 10\% galaxies and galaxy clusters.
For RASS sources contained
in the Bright Source Catalogue (BSC; Voges et al. 1996)
the identifications are compiled in the  Hamburg/RASS Catalogue of Optical 
Identifications (HRC) (Bade et al. 1998). The newest version
contains 4665 entries and is available on the web 
(http://www.hs.uni-hamburg.de/rass.html). The HRC provides plausible optical
candidates for 80\% of the BSC-RASS sources contained.

\section{ROSAC: A ROSAT based Search for AGN Clusters}
In the extragalactic sky about half of the RASS sources are AGN. 
Following the redshift distribution of RASS AGN given by Bade et al. (1995) 
70\% have redshifts z$<$0.5, giving a surface density of 0.5 AGN/deg$^2$.
Such a high surface density has not been reached by any survey before in this 
redshift range. Even the largest optical survey, the LBQS 
(Hewett et al. 1995), reached only 0.3 AGN/deg$^{2}$ 
(0.1$\le$z$\le$0.5) over an area of 454 deg$^{2}$. 
Presently the use of the RASS to study large-scale structure is however
 limited by the lack of redshift information for most of the AGN candidates.

For a first test, Tesch \& Engels (1998)  examined the spatial distribution 
of 856 RASS AGN,  for which  redshifts were available from the literature or 
from own follow-up observations. The AGN were compiled from an area of 
\mbox{$\sim$7000 deg$^2$,}
 in which optical identifications of RASS sources were made then.  
To analyze the data two different methods were used: a direct search for 
structures with the minimal spanning tree and a statistical study by the 
two-point correlation function.
The application of the minimal spanning tree led to the discovery of an 
AGN group with 7 members in a volume \mbox{V $\sim$ 140 $\times$ 75 $\times$ 
75 h$^{-3}$Mpc$^{3}$} in the Pisces constellation. With a mean redshift
z=0.27 this group is only the third discovered group at redshifts z$<$0.5. 
The two-point correlation function shows
AGN clustering on small scales ($\sim$10 h$^{-1}$Mpc) with a significance
$>4\sigma$, corroborating earlier results based on smaller AGN samples.

This first sample was however highly inhomogeneous. The area was defined
by the status of the identification process and included HQS survey fields
with very different amounts of X-ray absorption, leading to strongly
varying numbers of AGN candidates per field. In addition the number of
AGN per field with known redshifts was also strongly inhomogeneous.

\begin{figure}
\resizebox{\hsize}{!}{\includegraphics{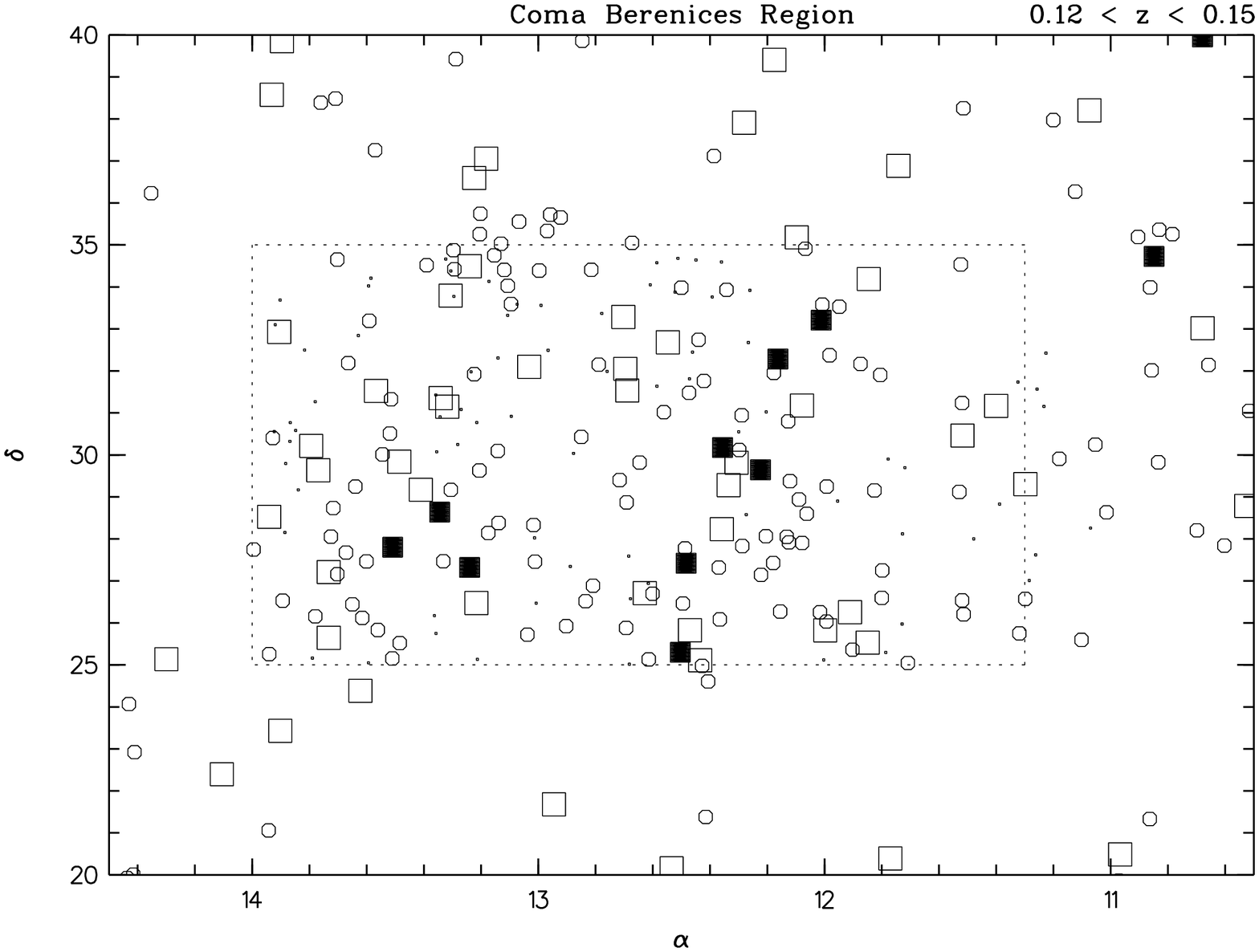}}
\caption[]{AGN distribution in the selected area Coma Berenices. A 
wall-like structure is outlined by the filled squares. Symbols as in
Figure 1}
\resizebox{\hsize}{!}{\includegraphics{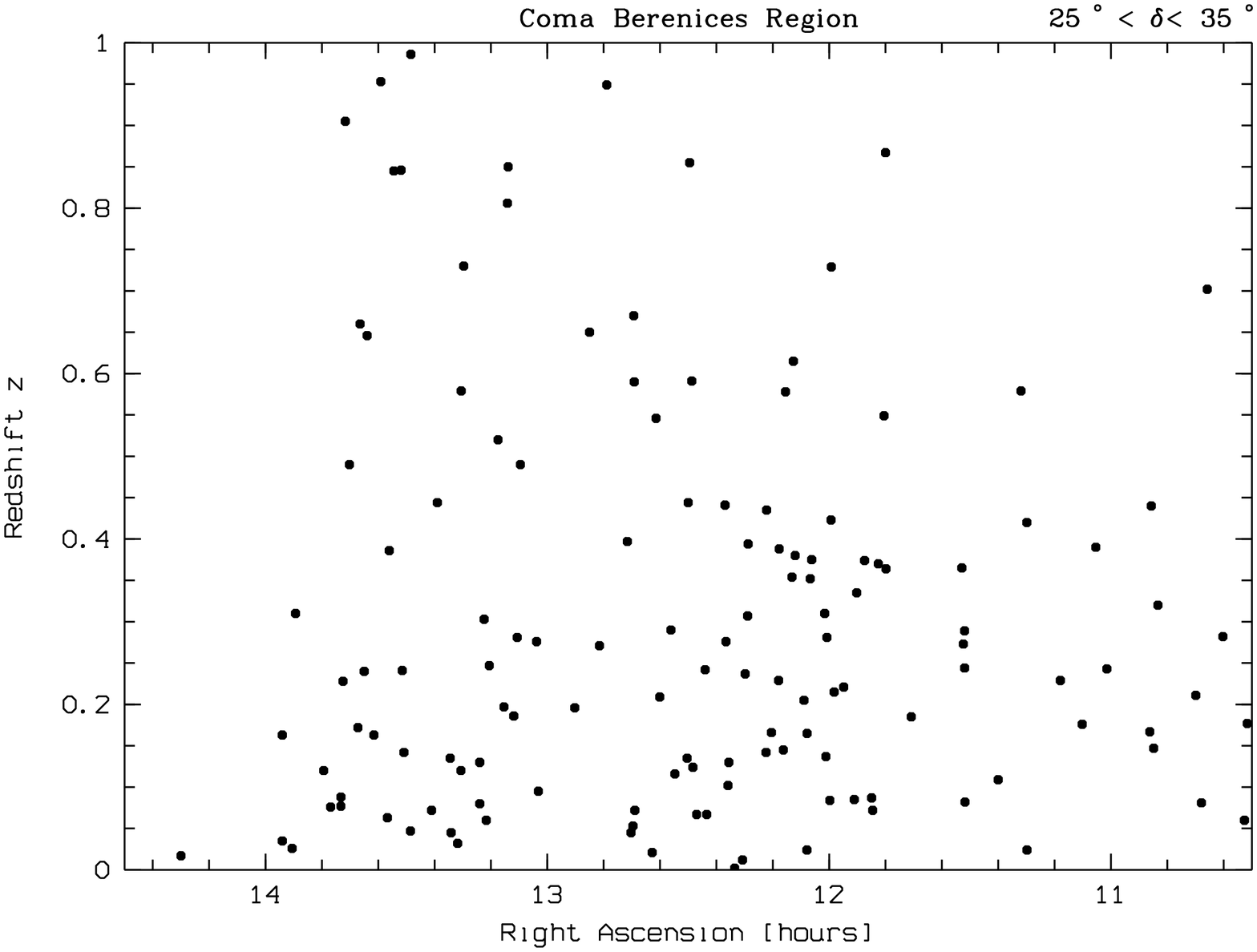}}
\caption[]{Selected area Coma Berenices. Slice merging all AGN
with 25$<\delta <$35$^{\circ}$ in the right ascension - redshift plane.}
\end{figure}

We started therefore a new program (ROSAC: A ROSAT based Search for
AGN Clusters; cf. http://www.hs.uni-hamburg.de/rosac.html) to obtain a
homogeneously selected sample with complete redshift information.
Three selected areas in the constellations Pisces, Ursa Major
and Coma Berenices were chosen. 
The main selection criteria for these areas have been a low
hydrogen column density N$_{H}$, a large number of already known redshifts
and the existence of interesting features as the new Pisces-AGN-Group or
further candidates of AGN groups with an insufficient significance level yet.
The cumulative area comprises $\approx$1700 deg$^{2}$ and contains 
$\approx$650 AGN and AGN candidates. 
The AGN surface densities are a function of the RASS detection limits, 
which depend mainly on N$_{H}$, and vary between 0.3 and 0.5 AGN/deg$^{2}$. 
In principle the surface densities can be increased further by inclusion
of the $\approx$40\% unidentified RASS sources in the Hamburg/RASS database.
This requires however additional CCD imaging and probably spectroscopy of 
more than one optical candidate to obtain the correct identification.
Unrecognized AGN are mostly not detected on the HQS objective prism plates
and are therefore optically rather weak (B$>$18.5).

Spectroscopy of RASS AGN candidates is underway using
telescopes on Calar Alto (2.2m), at Xinglong Station (2.16m), in 
Haute-Provence (1.93m), and at the Special Astrophysical Observatory (6m).
The confirmation rate is maintained at the 90\% level including BL Lac
objects. They make up $<$10\% of the AGN candidates and the S/N of their
spectra is usually insufficient to determine the redshift. 
For $\approx$70\% of the whole sample redshifts are available now.

\section{First results}
In all three regions we find in certain redshift ranges structures
which resemble clusters, walls and filaments as they were found in galaxy
distributions. This is exemplified in Figures 1 to 4 showing the
AGN distributions in the Ursa Major und the Coma Berenices regions.
For example, in the UMa region a filament extending $>75 h^{-1}$\,Mpc at
z$=0.105\pm0.025$ was found, while in the Coma region a wall-like
structure at z$=0.135\pm0.015$ was detected. 
At  these redshifts the probability of finding structures is 
relatively high since the redshift distribution of the RASS AGN peaks 
at z $\approx$ 0.1. The filament has a
size of $\approx$130 h$^{-1}$Mpc in redshift space, while the wall has
an overall size of 10 $\times$ 50 $\times$ 70 h$^{-3}$Mpc$^3$. 
Plotting redshifts against equatorial coordinates
the presence of voids is indicated, although their reality depends
strongly on the achieved completeness in spectroscopy. Completeness
is already 
achieved in part of the UMa region, in which a void is seen for example at
$\alpha$ = 9h 45m and z=0.15. Until now no statististical means have
been applied to these samples, leaving the reality of the detected structures
open. On larger scales, the AGN distributions resemble presently early maps 
of the spatial distribution of galaxies, as presented for example by
Joeveer \& Einasto (1978).  

Notwithstanding the confirmation of the detected structures their relation
to the matter distribution in the universe in general is open to debate.
AGN groups detected so far have relatively high redshifts making a direct
comparison to the distribution of galaxies or clusters of galaxies impossible.
Komberg et al. (1996) propose that the AGN groups evolve into the 
superclusters of galaxies in the local universe. Applying their
spatial number density of superclusters of 
n$_{\rm{SCL}} \approx$ 1.4$\cdot$\,10$^{-7}$ h$^3$ Mpc$^{-3}$
to the volume sampled by RASS AGN, Tesch \& Engels (1998) expected several
dozens AGN groups to be present, while they found only one. The inhomogeneity
of their sample precluded however a firm rejection of the Komberg et al.
hypothesis. The ROSAC sample will give stronger constraints.

The filaments and walls in the AGN distributions found in our selected
regions may indicate that the AGN and quasar groups are not isolated but
part of supercluster-void networks as they were discovered by
Einasto et al. (1997a,b) in the distribution of rich clusters of galaxies.
The non-detection of structures $>$20 h$^{-1}$ Mpc in the clustering
analyses of large optically selected quasar samples and the general
view that quasars avoid dense environments of galaxies (Smith et al. 1995)
is presently not consistent with a scenario in which the AGN distribution
trace the denser parts of the matter distribution in the universe.
AGN structures at low redshifts offer now the possibility to study
their relation to the galaxy distribution directly, as the
redshift determination
for galaxies at such distances is within the capabilities of the
modern optical telescopes.

\begin{acknowledgements}
We gratefully acknowledge support for the RASS identifications by grants
Re 353/22-1 to 4 of the Deutsche Forschungsgemeinschaft (DFG) and grant
DARA 50 OR 96016 of the BMBF. ROSAC is supported by the DFG under
grant En 176/13.
\end{acknowledgements}

\end{document}